\documentclass[iop, apjl]{emulateapj}
\usepackage{amsmath}

\usepackage{color,hyperref}

\shorttitle{Planets on the Edge}
\shortauthors{Valsecchi \& Rasio}

\begin{document}


\title{Planets on the Edge}


\author{Francesca Valsecchi\altaffilmark{1,2}, Frederic A. Rasio\altaffilmark{1,2}}
\affil{$^{1}$Center for Interdisciplinary Exploration and Research in
  Astrophysics (CIERA)} 
\affil{$^{2}$Department of Physics and Astronomy,
  Northwestern University, 2145 Sheridan Road, Evanston, IL 60208,
  USA.}  \email{francesca@u.northwestern.edu}


\begin{abstract}
Hot Jupiters formed through circularization of high-eccentricity orbits should be found at orbital separations $a$ exceeding {\it twice} that of their Roche limit $a_{\rm R}$. Nevertheless, about a dozen giant planets have now been found well within this limit ($a_{\rm R}\,\textless\,a\,\textless\,2\,a_{\rm R}$), with one coming as close as 1.2\,$a_{\rm R}$. In this Letter, we show that orbital decay (starting beyond 2\,$a_{\rm R}$) driven by tidal dissipation in the star can naturally explain these objects. For a few systems (WASP-4 and 19), this explanation requires the linear reduction in convective tidal dissipation proposed originally by \cite{Zahn1966, Zahn1989} and verified by recent numerical simulations \citep{PenevSRD2007}, but rules out the quadratic prescription proposed by \cite{GoldreichNicholson1977}. Additionally, we find that WASP-19-like systems could potentially provide direct empirical constraints on tidal dissipation,
as we could soon be able to measure their orbital decay through high precision transit timing measurements. 
%prescription through high precision transit timing measurements of their orbital decay rate. 
%{\bf prescription by observing their orbital decay through high precision transit timing measurements.}
\end{abstract}

\section{Introduction} \label{Intro}
Almost 200 of the known transiting exoplanets are giant planets with orbital periods less than 10\,days. 
These so-called hot Jupiters were most likely formed farther out at several AUs, but the debate continues on whether their tight orbits are the result of quasi-circular {\it disk migration} or {\it high-eccentricity migration}.
The first scenario involves slow orbital decay in a protoplanetary disk \citep{GoldreichTremaine80,Lin+96,Ward97,MurrayHHT98}, while the second involves tidal circularization of an orbit made extremely eccentric by gravitational interaction with companion stars, or between several planets \citep{RasioFord96,WuMurray03,FabryckyTremaine07,Nagasawa08,WuLithwick11,Naoz+11,PlavchanBilinski13}. 
In this letter, we focus on the hot Jupiters close to their Roche limits and show that they provide an important test for giant planet formation theories. In particular, this population provides constraints on the efficiency of convective damping of equilibrium tides (\citealt{Zahn1966,Zahn1989,GoldreichNicholson1977}; see also \citealt{Sasselov2003}).

In the disk-migration scenario, gas giants should be naturally found distributed in orbital separations, all the way down to the Roche limit $a_{\rm R}$. Instead, in any high-eccentricity migration scenario,  \cite{FordRasio06} pointed out that tidal circularization would lead to an inner edge at 2\,$a_{\rm R}$. While the great majority of systems are indeed observed to lie beyond 2\,$a_{\rm R}$ (e.g., \citealt{MatsumuraPR2010} and Fig.~\ref{fig:RocheLimit_planetMass} here), several hot Jupiters have now been discovered inside this limit. 
In a recent paper (\citealt{ValsecchiR+14}, hereafter VR14), we targeted giant planets in misaligned systems (where the stellar spin and orbital angular momentum are misaligned on the plane of the sky) and we showed that their properties could be naturally explained through high-eccentricity migration. Hot Jupiters can then be formed with a broad distribution of misalignments $\lambda$ and in orbits with a high eccentricity, which is quickly dissipated by planetary tides \citep{Jackson+08,MatsumuraPR2010}. Subsequently, stellar tides, magnetic braking, and stellar evolution lead to the observed distribution of $\lambda$ found around stars of different temperatures \citep{WinnFAJ10,Albrecht+12}. 
Here we consider the known hot Jupiters close to tidal disruption and investigate the possibility that these same physical mechanisms are responsible for bringing them inward from {\it beyond} 2\,$a_{\rm R}$. As the tides exerted on the star by the planet are expected to be too weak to keep up with the spin-down driven by magnetic braking \citep{BarkerOgilvie2009}, the resulting dissipation in the slowly rotating host star drives further orbital decay. 
%While it seems clear that this possibility might depend on the tidal prescription adopted, we find that half of the systems targeted in this work are consistent with this picture, independently on the tidal recipe. On the other hand, the remaining systems could be consistent with both disk- and high-eccentricity migration, depending on their current properties, as well as the assumed efficiency with which stellar tides are dissipated.  
With future measurements of the shift in transit times (e.g., \citealt{Sasselov2003,Birkby+14}), the orbital decay rate 
could be determined.
%If future measurements of the shift in transit arrival time will be available (e.g., \citealt{Sasselov2003,Birkby+14}), the orbital decay rate could be determined. 
This, in turn, would provide important constraints for both tidal dissipation and hot Jupiter formation theories.
In contrast to previous studies on these objects (e.g., \citealt{Sasselov2003,Gillon+14,Birkby+14}), we use detailed stellar models and compute the orbital evolution of hot Jupiters by integrating the equations describing the coupled evolution of the orbital elements and stellar spin (VR14).

This paper is organized as follows. We describe our hot Jupiter sample in \S~\ref{The Sample} and explain how we model each host star in \S~\ref{The Stellar Model}. In \S~\ref{The Orbital Evolution} we summarize the physical mechanisms included in our orbital evolution calculations and we emphasize the tidal prescriptions considered. We present our results in \S~\ref{Results} and conclude in \S~\ref{Discussion}.

In what follows $M_{*}$, $R_{*}$, $T_{\rm eff}$, Fe/H (or $Z$), $\lambda$ ($\Theta_{*}$), and $v_{{\rm rot}}{\rm sin}~i_{*}$ indicate the stellar mass, radius, effective temperature, metallicity, sky-projected (true) misalignment, and rotational velocity, respectively. The angles $\lambda$ and $\Theta_{*}$ are related via ${\rm cos}\,\Theta_{*} = {\rm sin\,}i_{*}\,{\rm cos\,}\lambda\,{\rm sin\,}i_{o}+{\rm cos\,}i_{*}\,{\rm cos\,}i_{o}$ \citep{FabryckyWinn09}. The angle between the stellar spin axis (the orbital angular momentum) and the line of sight is $i_{*}$ ($i_{o}$). The planetary mass (radius) is $M_{\rm pl}$ ($R_{\rm pl}$). The stellar spin frequency and planetary orbital frequency are $\Omega_{*}$ and $\Omega_{o}$, respectively. The orbital period (separation) is $P_{\rm orb}$ ($a$). 
%%%%%%%%%%%%%%%%%%%%%%%%%%%%%%%%%%%%%%%%%%%%%%%%
\section{Hot Jupiters Within 2$\,\lowercase{a}_{\rm R}$}\label{The Sample}
\begin{figure} [!h]
\epsscale{1.1}
\plotone{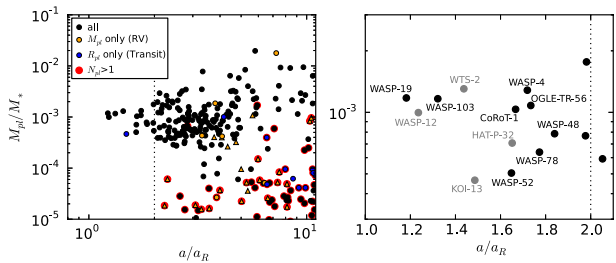}
\caption{Ratio of planetary to stellar mass as a function of the orbital distance to Roche limit ratio. The vertical dotted line is at $a/a_{\rm R}\,=\,2$. Left: exoplanets for which $P_{\rm orb}$, $M_{\rm *}$, and $M_{\rm pl}$ {\it or} $R_{\rm pl}$ are known within the range of $M_{\rm pl}/M_{*}$ and $a/a_{\rm R}$ displayed. In black are the systems where $P_{\rm orb}$, $M_{\rm *}$, $M_{\rm pl}$, and $R_{\rm pl}$ have been constrained. In blue (orange) are the systems where only $R_{\rm pl}$ ($M_{\rm pl}$) is known through transit [radial velocity (RV)] measurements. Like  \cite{MatsumuraPR2010}, when $R_{\rm pl}$ is unknown, we assume a Neptune (Jupiter) radius if $M_{\rm pl}\,\textless\,0.1\,M_{\rm J}$ ($\textgreater\,0.1\,M_{\rm J}$); when $M_{\rm pl}$ is unknown we assume a Neptune (Jupiter) mass if $R_{\rm pl}\,\textless\,0.35\,R_{\rm J}$ ($\textgreater\,0.35\,R_{\rm J}$). The orange triangles denote lower limits (only $M_{\rm pl}{\,\rm sin\,}i_{o}$ is known), while the red circles denote systems with multiple planets. Right: systems considered in this work (black) and systems where additional bodies have been found (in grey; \citealt{Santerne+12,Bechter+13,Knutson+13,Birkby+14}). The properties of the systems considered here have been updated from exoplanet.eu and we also include the recently discovered WTS-2 \citep{Birkby+14}. The size of the data points does not represent the uncertainties.}
\label{fig:RocheLimit_planetMass}
\end{figure}

%%%%%%%%%%%%%%%%%%%%%%%%%%%%%%%%%%%%%%%
The systems were queried from the NASA Exoplanet Archive on 20 February 2014. In Fig.~\ref{fig:RocheLimit_planetMass} we show $M_{\rm pl}/M_{*}$ as a function of $a/a_{\rm R}$ for the full sample of exoplanets currently known (left) and for the systems considered here (right). We adopt \citeauthor{Paczynski71}'s (\citeyear{Paczynski71}) approximation $a_{\rm R}\,=\,R_{\rm pl}/(0.462 q^{1/3})$, where $q\,=\,M_{\rm pl}/M_{*}\,\ll\,1$. The vertical dotted line marks the $a/a_{\rm R}\,=\,2$ limit, beyond which lie the great majority of systems. Here we focus on the hot Jupiters inside this limit where no additional bodies have been found (marked in grey, as such bodies could perturb the orbital evolution of the inner planet). We summarize their properties in Table~\ref{Tab:systemsParams}.
%%%%%%%%%%%%%%%%%%%%%%%%%%
%%%%%%%%%%%%%%%%%%%%%%%%%%
%Below, we summarize the parameters of each systems relevant to our analysis.
%For WASP-19 we adopt a stellar and planetary mass of $M_{*}\,=\,0.904\,\pm\,0.045\,M_{\odot}$           
%$M_{\rm pl}\,=\,1.114\,\pm\,0.040\,M_{\rm J}$, respectively. The stellar and planetary radius are $R_{*}\,=\,1.004\,\pm\,0.018\,R_{\odot}$ and $R_{\rm pl}\,=\,1.395\,\pm\,0.025\,R_{\rm J}$, respectively, and the orbital separation is $a\,=\,0.01616\,\pm\,0.00027\,$AU \citep{TR+13}. The stellar effective temperature and metallicity are $T_{\rm eff}\,=\,5440\,\pm\,60\,$K \citep{Maxted+11} and $Fe/H\,=\,0.02\,\pm\,0.09$, respectively \citep{Hebb+10}. For WASP-76 the same parameters are $1.46\,\pm\,0.07\,M_{\odot}$, $0.92\,\pm\,0.03\,M_{\rm J}$, $1.73\,\pm\,0.04\,R_{\odot}$, $1.83^{+0.06}_{-0.04}\,R_{\rm J}$, $0.0330\,\pm\,0.0005\,$AU, $6250\,\pm\,100\,$K, and $Fe/H\,=\,0.23\,\pm\,0.1$ \citep{West+13}. For WASP-103, \cite{Gillon+14} reports $1.220^{+0.039}_{-0.036}\,M_{\odot}$           
%$1.490\,\pm\,0.088\,M_{\rm J}$, $1.436^{+0.052}_{-0.031}\,R_{\odot}$, 
%$1.528^{+0.073}_{-0.047}\,R_{\rm J}$, $0.01985\,\pm\,0.00021\,$AU, $6110\,\pm\,160\,$K, and $Fe/H\,=\,0.06\,\pm\,0.13$. 
%Finally, for WTS-2  \cite{Birkby+13} determined $0.820\,\pm\,0.082\,M_{\odot}$           
%$1.12\,\pm\,0.13\,M_{\rm J}$, $0.761\,\pm\,0.033\,R_{\odot}$, $1.300\,\pm\,0.058\,R_{\rm J}$, $0.01855\,\pm\,0.00062\,$AU, $5000\,\pm\,250\,$K, and $Fe/H\,=\,0.2^{+0.3}_{-0.2}$. 
%%%%%%%%%%%%%%%%%%%%%%%%%%
%%%%%%%%%%%%%%%%%%%%%%%%%%
\section{Stellar Models}\label{The Stellar Model}
\begin{figure} [!h]
\epsscale{1.0}
\plotone{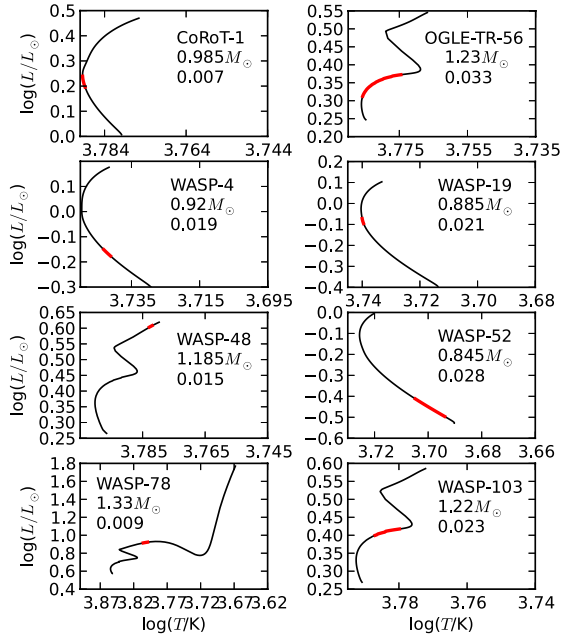}
\caption{Hertzsprung-Russell (HR) diagrams for all systems. The numbers quoted in each panel are the stellar initial mass and metallicity $Z$. The red solid lines represent the models which match the observed $Z$, $M_{*}$, $R_{*}$, and $T_{\rm eff *}$ within 1$\sigma$ (Table~\ref{Tab:systemsParams}), except for WASP-78 (where we can only match the observed $R_{*}$, and $T_{\rm eff *}$ within 2\,$\sigma$; see \S~\ref{The Stellar Model}).}
\label{fig:HR_all}
\end{figure}

The host star models are shown in Fig.~\ref{fig:HR_all} and are chosen from the grid of evolutionary tracks described in VR14 (computed with MESA; \citealt{PBDHLT2011,Paxton+13}) as follows. 
To be within the 1$\,\sigma$ uncertainties in $M_{*}$ and $Fe/H$, while still close to the mean observed values, OGLE-TR-56, WASP-4, 19, and 48's models are chosen randomly among those whose $M_{*}$ (Fe/H) are within 0.04\,$M_{\odot}$ (0.05) from the observed mean values at some point during the stellar evolution. The same procedure is applied to CoRoT-1 and WASP-103, but requiring the limit on $M_{*}$ (Fe/H) to be 0.03\,$M_{\odot}$ (0.05). 
%We chose a narrower range for $M_{*}$ because it is generally measured more precisely. 
The age of WASP-52 varies by several Gyr, depending on its properties. 
\cite{Hebrard+13} report a lower limit of 0.5 Gyr from lithium abundance and quote a gyrochronological age of 0.4$^{+0.3}_{-0.2}\,$Gyr derived from the observed $v_{\rm rot}{\rm sin\,}i_{*}$ (\citealt{Hebrard+13} and references therein). However, $i_{*}$ is not known.
To see some degree of orbital evolution we take the model that reaches the oldest age within the $1\,\sigma$ uncertainties in $M_{*}$ and $Fe/H$. This model spans a range $\simeq1.5\,-\,7\,$Gyr. Finally, WASP-78 is not in VR14's catalogue, and we evolve a star with the observed mean $M_{*}$ and $Z$. The observed $R_{*}$ and $T_{\rm eff}$ can be matched only within 2$\,\sigma$. The agreement could be improved by varying some of the physics entering the stellar modeling (e.g., the mixing length parameter, which is usually varied between 1$-$2 in the literature, see \citealt{PBDHLT2011, Paxton+13}). However, we choose not to introduce additional free parameters and therefore apply the same physical assumptions to all systems.

%%%%%%%%%%%%%%%%%%%%%%%%%%
\section{Orbital Evolution}\label{The Orbital Evolution}

The procedure and assumptions adopted in our calculations are explained in detail in VR14. Here we outline the main points for clarity and present the new tidal prescriptions considered. 
We study CoRoT-1\,b, OGLE-TR-56\,b, WASP-4, 19, 48, 52, and 103\,b's evolutionary past, by scanning the initial [at the stellar Zero Age Main Sequence (ZAMS)] parameter space made of $P_{\rm orb}$, $\Omega_{*}/\Omega_{o}$, and $\Theta_{*}$ (when measured). We then integrate the equations describing the coupled evolution of $a$, $\Omega_{*}$, and $\Theta_{*}$, due to stellar tides, wind mass loss, magnetic braking, and the evolution of the host star (\S~\ref{The Stellar Model}). 
All orbits are consistent with circular and we assume that damping of the eccentricity occurred quickly through dissipation in the planet (this assumption is discussed in VR14). 
For all parameters described below, we adopt the same values used in VR14, unless stated otherwise.

For stellar wind mass loss and magnetic braking we proceed as in VR14, introducing the parameter $\gamma_{\rm MB}$, which controls the strength of angular momentum loss via magnetic braking. 
%entering the magnetic braking prescription, which we vary between 0$\,-\,$1, thus bracketing the typical values considered in the literature for the stellar types of interest here (e.g.,  \citealt{BarkerOgilvie2009,MatsumuraPR2010}). 
%%%%%%%%%%%%%%%%%%%%%%%%%%
For tides, we use the weak-friction approximation when there is no information about misalignment (and we then take $\Theta_{*}\,=\,0$). Instead, for systems where the misalignment has been constrained, we include the effect of inertial wave dissipation (IWD) following \cite{Lai12}, and consider a variety of initial $\Theta_{*}$.
%, to explain the degree of  misalignments found around stars of different temperatures.
When accounting for weak-friction tides {\it alone}, we consider both sub- and super-synchronous initial configurations ($\Omega_{*}\textless\Omega_{o}$ and $\Omega_{*}\textgreater\Omega_{o}$, respectively) and vary $\Omega_{*}/\Omega_{o}$ between 0 and 100 in steps of 0.2. We halt the calculation if the star is spinning faster than break-up. Instead, when accounting for IWD, we consider initial $\Omega_{*}/\Omega_{o}$ values only up to 0.9, according to the validity of the \cite{Lai12} prescription. Furthermore, we consider different values for the efficiency of IWD by varying the tidal quality factor $Q'_{\rm 10}$.

%Here we note that the stellar inclination is unconstrained for the systems studied in detailed here (apart for WASP-19 where the observations suggest  that the stellar spin and orbital angular momentum are close to alignment, see \citealt{TR+13}). Therefore, we draw a random uniform distribution in cos$\,i_{*}$ between 0\,-\,1 and consider 50 values of cos$\,i_{*}$ for each combination of initial orbital and stellar spin configuration. 
%To summarize, for systems with no misalignment information we scan on initial $P_{\rm orb}$, $\Omega_{*}/\Omega_{o}$, and $\gamma_{\rm MB}$. Instead, for systems with an observed $\lambda$, we scan on initial $P_{\rm orb}$, $\Omega_{*}/\Omega_{o}$, $\Theta_{*}$, and $\gamma_{\rm MB}$, for the four different values of $Q'_{\rm 10}$ mentioned above. 
Thus far, we followed VR14, apart for enforcing the validity of the \citeauthor{Lai12}'s (\citeyear{Lai12}) recipe for tides (initial $\Omega_{*}/\Omega_{o}\textless\,1$). Now we go one step further and vary the weak-friction tides prescription following \cite{Sasselov2003}. In VR14 we included the effects of both convective damping of the equilibrium tide and radiative damping of the dynamical tide and showed that, in the weak-friction regime, the former always dominates for typical hot Jupiter systems. For convective dissipation, we followed the mixing-length theory of convection and assumed that the oscillatory tidal distortion is dissipated by turbulent (eddy) viscosity. For high tidal forcing frequencies, the efficiency of angular momentum transport by the largest eddies is inhibited, and the exact form of this reduction is still under debate. 
Defining the reduction factor as 
\begin{align}
&f_{\rm *,conv} = min\left[1, \left(\frac{P_{\rm *,tid}}{2\tau_{\rm *,conv}}\right)^s\right],\label{eq:fConv}
\end{align}
where $P_{\rm *,tid}$ and $\tau_{\rm *,conv}$ are the tidal forcing period and the convective turnover timescale (VR14 and references therein), 
two commonly used prescriptions are the linear one ($s\,=\,1$) proposed by \cite{Zahn1966,Zahn1989} and supported by \citeauthor{PenevSRD2007}'s (\citeyear{PenevSRD2007}) recent numerical simulations, and the quadratic one ($s\,=\,2$) proposed by \cite{GoldreichNicholson1977}. 
Here we follow \cite{Sasselov2003} and consider both $s\,=\,1$ and $s\,=\,2$ (hereafter the {\it Zahn} and {\it GN} prescription, respectively).
%\begin{equation}
%P_{\rm *,tid}and we adopted $s\,=\,2$, following  \cite{GoldreichNicholson1977}. In Eq.~(\ref{eq:fConv}), $\tau_{\rm *,conv}$ is the convective turnover timescale, while the tidal forcing period is given by
%\begin{equation}
%P_{\rm *,tid} = \left|\frac{1}{P_{\rm orb}} - \frac{1}{P_{\rm *, spin}}\right|^{-1}
%\end{equation}
%where $P_{\rm *, spin}$ is the stellar spin period

At each time-step during the calculation we check that the planet is within its Roche lobe \citep{Paczynski71}. We stop the evolution when the model's $M_{*}$, $R_{*}$, $T_{\rm eff,*}$, $v_{\rm rot}{\rm sin\,}i_{*}$, and $\Theta_{*}$ (if constrained) are within 1$\,\sigma$ (2$\,\sigma$ in $R_{*}$ and $T_{\rm eff,*}$ for WASP-78) from the observed values and $P_{\rm orb}$ crosses the present value (Table~\ref{Tab:systemsParams}). 
%%%%%%%%%%%%%%%%%%%%%%%%%%%%%%%%%%%%%%%%%%%%%%%%

\section{Results} \label{Results}

Our numerical results are summarized in Table~\ref{Tab:orbiEvolResults}. 
The parameter $T_{\rm shift}$, is the transit arrival time shift, which we computed following \S~7.2 of \cite{Birkby+14} and our $\dot{a}$  values. 
Here we follow \cite{Birkby+14} and assume 10 years of observations with a timing accuracy of 5\,s \citep{Gillon+09}. However, note that orbital periods are routinely measured to less than 1\,s with multiple observations (Jason Steffen, private communication). In Table~\ref{Tab:orbiEvolResults}, we therefore list the full range of $T_{\rm shift}$ values computed.
%and list $T_{\rm shift}$ when $\textgreater\,5\,$s. We only provide an order of magnitude estimate when $T_{\rm shift}\,\textless\,5\,$s.
We perform a first scan in initial orbital periods ($P_{\rm orb, in}$) with a coarse resolution, which we then increase during a second scan, if needed. In Table~\ref{Tab:orbiEvolResults} we list the results with limited precision, just to give a sense for the possible initial orbital configurations. 

%We perform a first scan in initial orbital perioda ($P_{\rm orb, in}$) with a coarse resolution, which we then increase during a second scan, if needed. For OGLE-TR-56, WASP-48, and 103 we consider $P_{\rm orb, in}$ between 0.5$\,-\,6\,$d, 2.5$\,-\,4.5\,$d, and 2$\,-\,4.5\,$d, in steps of 0.2, 0.1, and 0.05, respectively.
%For CoRoT-1 and WASP-4 we consider $P_{\rm orb, in}$ between 2$\,-\,5\,$d in steps of 0.2\,d and 0.1\,d, respectively, for $s$\,=\,1, and 1$\,-\,5\,$d and 1.5$\,-\,3.5\,$d, respectively, with the same step sizes for $s\,=\,2$. For WASP-19 and 52 we consider $P_{\rm orb, in}$ between 3$\,-\,5\,$d and 1.5$\,-\,3.5\,$d, respectively, for $s\,=\,1$, and between 1$\,-\,3\,$d and 1.5$\,-\,2.5\,$d, respectively, for $s\,=\,2$, in steps of 0.05\,d. Finally, for WASP-78 we consider $P_{\rm orb, in}$ between 1.5$\,-\,6\,$d in steps of 0.1\,d. In Table~\ref{Tab:orbiEvolResults} we list the results with limited precision, just to give a sense for the possible initial orbital configurations. 

Half of the systems considered are easily explained: CoRoT-1, OGLE-TR-56, WASP-48, and 103, according to our detailed modeling, started their orbital evolution from beyond 2\,$a_{\rm R}$, independent of the tidal prescription adopted ({\it Zahn} or {\it GN}). On the other hand, the evolutionary picture differs for WASP-4 and 19, where only the {\it Zahn}  prescription is consistent with $a_{\rm in}\,\textgreater\,2\,a_{\rm R}$. While for WASP-4 $T_{\rm shift}$ is lower than the 5\,s limit considered by \cite{Birkby+14}, it is 5$-$8 times this limit for WASP-19, according to {\it Zahn}. If detected, {\bf it} could provide important constraints on tidal dissipation theory. Interestingly, WASP-19-like systems could also be used to constraint the efficiency of IWD. In fact, within the range of $Q'_{\rm 10}$ values considered and given the resolution of our initial parameter space, this system can be explained by the {\it Zahn} ({\it GN}) prescription only if $Q'_{\rm 10}\,\textless\,10^{10}$ ($Q'_{\rm 10}\,\geq\,10^{7}$). 
%Of course we can not exclude that this result is due to the resolution adopted during the scan of the initial parameter space. However, a detailed study like the one presented here on the full sample of hot Jupiters in misaligned systems could reveal interesting trends between $Q'_{\rm 10}$ and the components properties or the orbital configuration.
Finally, WASP-52 and 78 are consistent with $a_{\rm in}\,\textgreater\,2\,a_{\rm R}$ only for the longest $P_{\rm orb,in}$ considered. For WASP-52, a more precise determination of its age could provide constraints on the more likely migration scenario. Furthermore, within our initial parameter space, this system cannot be explained by the {\it GN} prescription. This is due to the upper limit imposed on the initial $\Omega_{*}/\Omega_{o}$.
For WASP-78 the efficiency of tides is never reduced and both tidal prescriptions predict the same evolutionary picture. This system could still be used to constrain hot Jupiter formation theories from measurements of $T_{\rm shift}$.

The ages constrained with our modeling (Table~\ref{Tab:orbiEvolResults}) agree with those reported in the literature (when available) for most systems. WASP-48's age is uncertain and, even though we list the one derived by the lack of lithium and Ca H+K, we note that the rotation rate supports an age of $0.6^{+0.4}_{-0.2}\,$Gyr. Alternatively, isochrones analysis yields an age of $3.0^{+1.0}_{-0.5}\,$Gyr (\citealt{Enoch+11b} and references therein). 

The parameter $\gamma_{\rm MB}$ is generally set to 0.1 and 1 for F- and G-dwarfs \citep{BarkerOgilvie2009,MatsumuraPR2010}, respectively. We find solutions for nearly any of the $\gamma_{\rm MB}$ values considered for OGLE-TR-56, WASP-48, and 52, with both tidal prescriptions, and for WASP-4, and 103 with the {\it Zahn} prescription. Furthermore, the $\gamma_{\rm MB}$ range that explains WASP-78 ($\gamma_{\rm MB}\,\leq\,0.3$) and 103 ($\gamma_{\rm MB}\,\leq\,1$ and 0.2 for {\it Zahn} and {\it GN}, respectively) encloses the value 0.1 generally adopted for F-dwarfs. Instead, CoRoT-1, WASP-4 , and 19, with G-dwarfs, all have $\gamma_{\rm MB}\,\textless\,1$. According to {\it Zahn} ({\it GN}), we find  $\gamma_{\rm MB}\,\leq\,0.1$ (0.2) both for CoRoT-1 and WASP-19. On the other hand, for WASP-4, the {\it GN} prescription allows $\gamma_{\rm MB}\,\leq\,0.2$ (0.9) for $Q'_{\rm 10}\,=\,10^{6}$ ($\geq\,10^{7}$). This discrepancy and the fact that the {\it GN} prescription can not explain WASP-52 is due to the upper limit on the initial $\Omega_{*}/\Omega_{o}$ considered here (0.9), as the \cite{Lai12} recipe for tides is strictly valid for sub-synchronous systems. In this regime, IWD affects only $\Omega_{*}$ and $\Theta_{*}$, while it might affect $a$ when $\Omega_{*}\,\geq\,\Omega_{o}$ (see VR14), but we do not account for this possibility. Since we find evolutionary solutions for initial $\Omega_{*}/\Omega_{o}$ values up to 0.9, super-synchronous configurations would likely yield more solutions and higher  $\gamma_{\rm MB}$ values. However, a different prescription for the evolution of $a$ should then be adopted or a detailed study of the significance of IWD compared to the other physical mechanisms should be performed (VR14). 
%For example, stretching the validity of the \citeauthor{Lai12}'s (\citeyear{Lai12}) prescription to include initial $\Omega_{*}/\Omega_{o}\,=\,1$, we find that WASP-52 could have started its evolution with $P_{\rm orb,in}\,\simeq\,$1.8\,d ($a/a_{\rm R}\,=\,1.7$), and that, if $Q'_{\rm 10}\,\geq\,10^{8}$, IWD timescales are at least one order of magnitude longer than the main driver of spin and misalignment evolution. This suggest that IWD might be inefficient in driving the evolution of $a$ if configurations where $\Omega_{*}/\Omega_{o}\,\textgreater\,1$ and $Q'_{\rm 10}\,\geq\,10^{8}$ are considered.

\begin{figure} [!h]
\epsscale{1.0}
\plotone{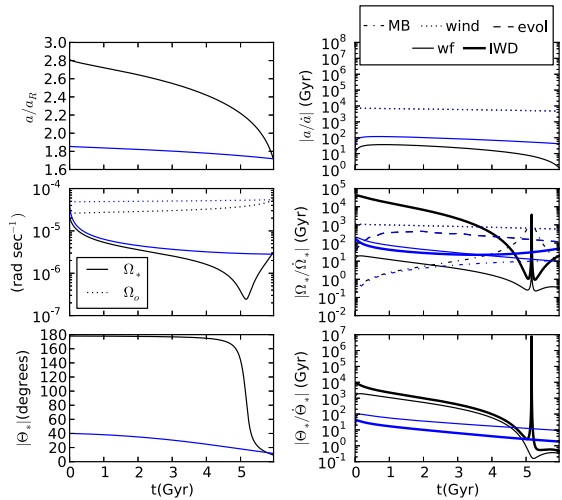}
\caption{Detailed orbital evolution of a WASP-4-like system. Left:  evolution of the orbital separation (top), stellar spin and orbital frequency (middle), and misalignment (bottom). Right:  evolution of the timescales associated with the physical effects considered. Specifically, ``wf'' refers to tides in the weak friction approximation [Eqs.~(1)-(3) in VR14], ``evol'' refers to changes in the stellar moment of inertia [Eqs.~(10) in VR14], ``IWD'' refers to dissipation of inertial waves [the sum of the second and third terms on the right side of Eqs.~(17) and (18) in VR14], ``wind'' refers to stellar wind mass loss [Eqs.~(11)-(12) in VR14], and ``MB'' refers to magnetic braking [Eqs.~(13) in VR14]. In black (blue) is the evolution according to {\it Zahn} ({\it GN}). For the {\it Zahn} ({\it GN}) example, the initial conditions are: $P_{\rm orb} = 2.8\,$d (1.5\,d), $\Omega_{*}/\Omega_{o}  = 0.8$ (0.8), $\Theta_{*} = 178^{o}$ (40$^{o}$), and $\gamma_{MB}\,=\,1.0$ (0.8). Furthermore, $Q_{10}'\,=\,10^{7}$ and $i_{*}\,=\,85^{o}$. As $i_{o}\simeq\,89^{o}$, the true and sky-projected misalignments are similar.}
\label{fig:DetailedOrbitalEv_WASP}
\end{figure}

In Fig.~\ref{fig:DetailedOrbitalEv_WASP} we show, as an example, the detailed evolution of a WASP-4-like system according to {\it Zahn} ({\it GN}) in black (blue). 
%This is an interesting system since it might have started its orbital evolution beyond 2$\,a_{\rm R}$ only with the {\it Zahn} prescription for tides.
This is an interesting system since only with the {\it Zahn} prescription could its orbit have begun beyond 2$\,a_{\rm R}$.
%%%%%%%%%%%%%%%%%%%%%%%%%%
%%%%%%%%%%%%%%%%%%%%%%%%%%%%%%%%%%%%%%%%%%%%%%%%

For both tidal prescriptions, the evolution of $a$ is driven by convective damping of the equilibrium tide, which drives to orbital decay (top panels). The evolution of $\Omega_{*}$ is driven overall by magnetic braking if the {\it GN} prescription is adopted. This causes the star to spin-down. Instead, according to {\it Zahn}, magnetic braking dominates for the first $\simeq\,2$\,Gyr and the remaining evolution is driven by weak-friction tides. The latter contributes to spin-down until the system is $\simeq\,5$\,Gyr old and $\Theta_{*}\textgreater\,90^{o}$. After $\Theta_{*}$ has crossed 90$^{o}$ (marked by a sudden peak in the timescales) the derivative describing the tidal evolution of the stellar spin in the weak-friction regime changes sign, and tides tend to synchronize $\Omega_{*}$ with $\Omega_{o}$. 
Finally, the evolution of the misalignment is driven by IWD with the {\it GN} prescription and by both weak-friction tides and IWD with the {\it Zahn} prescription. These effects cause the misalignment to decrease to the currently observed value.
%%%%%%%%%%%%%%%%%%%%%%%%%%%%%%%%%%%%%%%%%%%%%%%%
\section{Conclusion} \label{Discussion}
We investigated tidal dissipation and giant planet formation theories, by focusing on hot Jupiters with orbits close to the Roche limit ($a_{\rm R}$). In particular, we tested whether their properties are consistent with high-eccentricity migration $-$ where the highly eccentric orbits of giant planets are tidally circularized, through tidal dissipation in the planet, to distances larger than $2a_{\rm R}$, and later orbital decay is produced by tidal dissipation in the star.
%We investigated tidal and giant planet formation theories {\bf by targeting the observed hot Jupiters whose orbits lie within twice the Roche limit ($2\,a_{\rm R}$)}. 
%FIXME
%In particular, we tested whether their properties are consistent with high-eccentricity migration, where hot Jupiters are brought close to their stellar host {\it beyond} 2$a_{\rm R}$ via tidal circularization of a highly eccentric orbit; during the subsequent evolution, tides in the star cause orbital decay all the way to {\it within} 2$\,a_{\rm R}$. 
%{\bf In particular, we tested whether their properties are consistent with high-eccentricity migration $-$ where the eccentric orbits of Jovian planets are tidally circularized to distances larger than 2$\,a_{\rm R}$$-$ followed by orbital decay from stellar tides.}
We studied CoRoT-1 b, OGLE-TR-56 b, WASP-4, 19, 48, 52, 78, and 103 b and computed the past evolution of their orbital separation, stellar spin, and misalignment (when observed), including the effects of stellar tides and wind mass loss, magnetic braking, and the evolution of the host star. For the reduction in the effectiveness of convective damping of the equilibrium tide when the forcing period is less than the turnover period of the largest eddies, we tested the linear and quadratic theory of \cite{Zahn1966,Zahn1989} and \cite{GoldreichNicholson1977}, respectively.

We found that CoRoT-1, OGLE-TR-56, WASP-48, and 103 are consistent with high-eccentricity migration, independent of the tidal prescription adopted. This same conclusion may hold for WASP-78, depending on its initial orbital configuration. This could be validated by future measurements of the transit arrival time shift ($T_{\rm shift}$, e.g., \citealt{Sasselov2003,Birkby+14}). 
Within the parameter space considered here, WASP-52 can only be explained by the \citeauthor{Zahn1966}'s (\citeyear{Zahn1966,Zahn1989}) prescription.
Furthermore, this system could be consistent with high-eccentricity migration, depending on its initial orbital configuration. While $T_{\rm shift}$ for WASP-52 might be too small to detect, a more precise determination of its age could be used to distinguish between the different migration scenarios.
Finally, WASP-4 and WASP-19 are consistent with high-eccentricity migration only according to \citeauthor{Zahn1966}'s (\citeyear{Zahn1966,Zahn1989}) prescription. For WASP-19 in particular, the fairly rapid orbital decay could lead to a significant $T_{\rm shift}$ which, if detected, would provide an important confirmation of these ideas.

The 3-D numerical simulations by \cite{PenevSRD2007} showed a reduction factor that closely matched the linear prescription by \cite{Zahn1966,Zahn1989}. With this prescription, the results presented here show that {\it all} systems currently known close to their Roche limit are indeed consistent with a high-eccentricity migration scenario for the formation of hot Jupiters. %However, future empirical determinations of orbital decay rate for the hot Jupiters closest to their Roche limit by precise timing of their transit will provide a powerful laboratory. These measurements combined with theoretical estimate of orbital decay rates for different tidal prescriptions will provide a testbed for both tidal dissipation and hot Jupiter formation theories. We urge the observers to make these measurements.
%------------------------------------------------------------------------------------------------------------------------------------------------------------------------
 \acknowledgments
%\appendix

\begin{acknowledgements}
 This work was supported by NASA Grant NNX12AI86G. We thank Jason Steffen for his constructive remarks and comments.
% Computational resources supporting this work were provided by the Northwestern University ÒGrailÓ cluster, purchased with a National Science Foundation Major Research Instrumentation award (PHY-1126812), and by the Northwestern University Quest High Performance Computing (HPC) cluster. 
This research has made use of the NASA Exoplanet Archive.
%, which is operated by the California Institute of Technology, under contract with the National Aeronautics and Space Administration under the Exoplanet Exploration Program.
\end{acknowledgements}

%\bibliography{myBibtex}{}
\bibliographystyle{apj}

%%%%%%%%%%%%%%%%%%%%%%%%%%
\begin{deluxetable}{lcccccccccc}
\centering
\tabletypesize{\scriptsize}
    \tablecaption{Systems relevant for this work.}
\tablehead{
\colhead{Name} & \colhead{$M_{\rm pl}$} & \colhead{$R_{\rm pl}$} & \colhead{$M_{*}$} &\colhead{$R_{*}$ } & \colhead{$T_{\rm eff,*}$} & \colhead{$Fe/H$}&
\colhead{$v_{\rm rot}{\rm sin\,}i_{*}$} & \colhead{$P_{\rm orb}$} & \colhead {$\lambda$}&\colhead {$i_{o}$}
\\
 & \colhead{$(M_{\rm J})$} & \colhead{$(R_{\rm J})$} & \colhead{$(M_{\odot})$} &\colhead{$(R_{\odot})$ } & \colhead{(K)} & &
\colhead{(km\,s$^{-1}$)} & \colhead{(d)} &\colhead{(deg)}&\colhead{(deg)}
}
\startdata
C-1$^{1}$ & 1.03 & 1.49 & 0.950$^{+0.150}_{-0.150}$ & 1.110$^{+0.050}_{-0.050}$ & 6298$^{+150}_{-150}$ & -0.30$^{+0.25}_{-0.25}$ & 5.20$^{+1.00}_{-1.00}$& 1.509 & $^{2}$77.0$^{+11.0}_{-11.0}$ & $^{3}$83.80 \\
\smallskip
O-56$^{4}$ & 1.39 & 1.36 & 1.228$^{+0.072}_{-0.078}$ & 1.363$^{+0.089}_{-0.086}$ & 6050$^{+100}_{-100}$ & 0.22$^{+0.10}_{-0.10}$ & $^{5}$3.20$^{+1.00}_{-1.00}$& 1.212 & $-$ & $^{6}$73.72 \\
\smallskip
W-4$^{7}$ & 1.24 & 1.41 & 0.920$^{+0.060}_{-0.060}$ & 0.907$^{+0.014}_{-0.013}$ & $^{8}$5500$^{+100}_{-100}$ & $^{8}$-0.03$^{+0.09}_{-0.09}$ & $^{9}$2.14$^{+0.37}_{-0.37}$& 1.338 & -1.0$^{+14.0}_{-12.0}$ & 88.80 \\
\smallskip
W-19$^{10}$ & 1.11 & 1.40 & 0.904$^{+0.045}_{-0.045}$ & 1.004$^{+0.018}_{-0.018}$ & $^{11}$5440$^{+60}_{-60}$ & $^{12}$0.02$^{+0.09}_{-0.09}$ & $^{*}$4.30$^{+0.15}_{-0.15}$& 0.789 & 1.0$^{+1.2}_{-1.2}$ & 78.94 \\
\smallskip
W-48$^{13}$ & 0.98 & 1.67 & 1.190$^{+0.050}_{-0.050}$ & 1.750$^{+0.090}_{-0.090}$ & 5920$^{+150}_{-150}$ & -0.12$^{+0.12}_{-0.12}$ & 3.20$^{+0.30}_{-0.30}$& 2.144 & $-$ & 80.09 \\
\smallskip
W-52$^{14}$ & 0.46 & 1.27 & 0.870$^{+0.030}_{-0.030}$ & 0.790$^{+0.020}_{-0.020}$ & 5000$^{+100}_{-100}$ & 0.03$^{+0.12}_{-0.12}$ & 2.50$^{+1.00}_{-1.00}$& 1.750 & 24.0$^{+17.0}_{-9.0}$ & 85.35 \\
\smallskip
W-78$^{15}$ & 0.89 & 1.70 & 1.330$^{+0.090}_{-0.090}$ & 2.200$^{+0.120}_{-0.120}$ & 6100$^{+150}_{-150}$ & -0.35$^{+0.14}_{-0.14}$ & 7.20$^{+0.80}_{-0.80}$& 2.175 & $-$ & 83.20 \\
\smallskip
W-103$^{16}$ & 1.49 & 1.53 & 1.220$^{+0.039}_{-0.036}$ & 1.436$^{+0.052}_{-0.031}$ & 6110$^{+160}_{-160}$ & 0.06$^{+0.13}_{-0.13}$ & 10.60$^{+0.90}_{-0.90}$& 0.926 & $-$ & 86.30 \\
\enddata
\tablerefs{Following \href{http://exoplanet.eu/catalog/}{exoplanet.eu};
$(1)$: \citealt{Barge2008};
$(2)$: \citealt{Pont+10};
$(3)$: \citealt{Borsa2013};
$(4)$: \citealt{Torres+08};
$(5)$: \citealt{Melo+06};
$(6)$: \citealt{Adams+11};
$(7)$: \citealt{SanchisO+11};
$(8)$: \citealt{Gillon+09};
$(9)$: \citealt{Triaud+10};
$(10)$: \citealt{TR+13};
$(11)$: \citealt{Maxted+11};
$(12)$: \citealt{Hebb+10};
$(13)$: \citealt{Enoch+11b};
$(14)$: \citealt{Hebrard+13};
$(15)$: \citealt{Smalley+12};
$(16)$: \citealt{Gillon+14}.}
\tablecomments{Observed properties of CoRoT-1 (C-1), OGLE-TR-56 (O-56), WASP-4 (W-4), WASP-19 (W-19), WAP-48 (W-48), WASP-52 (W-52), WASP-78 (W-78), and WASP-103 (W-103). The symbols are defined in \S~\ref{Intro}. The parameters listed with limited accuracy are those for which we use only the mean value. 
%Assuming $Z_{\odot}\,=\,0.02$, the stellar models representative of  C-1, O-56, W-4, W-19, W-48, W-52, W-78, and W-103 have an initial mass ($Fe/H$) of 0.985$M_{\odot}$ (-0.46), 1.23$M_{\odot}$ (0.22), 0.92$M_{\odot}$ (-0.02), 0.885$M_{\odot}$ (0.02), 1.185$M_{\odot}$ (-0.12), 0.865$M_{\odot}$ (0.02), 1.315$M_{\odot}$ (-0.26), and 1.22$M_{\odot}$ (0.06), respectively, and were chosen from the grid of evolutionary tracks described in VR14 as follows. 
%To be within the 1$\,\sigma$ uncertainties, while still close to the mean observed values, the models representative of O-56, W-4, W-19, W-48, and W-78 were chosen randomly among those whose $M_{*}$ (Fe/H) are within 0.04\,$M_{\odot}$ (0.05) from the observed mean values at some point during the stellar evolution. This same procedure is applied on W-52, W-103, and C-1,  requiring these same limits on $M_{*}$ (Fe/H) to be 0.03\,$M_{\odot}$ (0.05) for the former two systems, and 0.04\,$M_{\odot}$ (0.2) for the latter. We chose to be closer to the stellar mass than metallicity, as $M_{*}$ is measured more precisely.
}
\tablenotetext{*}{The rotational velocity for WASP-19 is the {\it true} equatorial velocity.}
%\tablenotetext{b}{Another sample footnote for table~\ref{tbl-1}}
\label{Tab:systemsParams}
\end{deluxetable}
%%%%%%%%%%%%%%%%%%%%%%%%%%
%%%%%%%%%%%%%%%%%%%%%%%%%%%%%%%%%%%%%%%%%%%%%%%%
\begin{deluxetable}{ccccccccccccc}
\tabletypesize{\scriptsize}
\tablecaption{Initial $a/a_{\rm R}$}
\tablehead{
  & & & && \colhead{\small{Zahn (GN)}}  &  &  & &   
\textcolor{white}{1}\\
%&&&&&{\bf Zahn} (GN)
\\
\hline
\textcolor{white}{1}\\
\colhead{Name}  &\colhead{$a_{\rm pr}$} &  \colhead{$t_{*,lit}$}&$Q'_{\rm 10}$& \colhead{$a_{\rm in}$}  & \colhead{$P_{\rm orb, in}$} &\colhead{$T_{\rm shift}$}& \colhead{$t_{*,mod}$} 
\\
&\colhead{($a_{\rm R}$)} & \colhead{(Gyr)} &&\colhead{($a_{\rm R}$)} & \colhead{(days)} & \colhead{(s)}& \colhead{(Gyr)} &}
\textcolor{white}{1}\\
\startdata
\textcolor{white}{1}\\
C-1 & 1.7 & & $10^{6}$& 2.7\,-\,3.1 (2.0\,-\,2.7) & 3.2\,-\,3.8 (2.0\,-\,3.2) & $\simeq$\,2.2$-$3.7 (0.3$-$0.8) & 4.3\,-\,5.1\\
       &        & & $\geq\,10^{7}$& 2.6\,-\,3.1 (2.0\,-\,2.7) & 3.0\,-\,3.8 (2.0\,-\,3.2) &  $\simeq$\,2.2$-$3.8 (0.3$-$0.8)  & 4.3\,-\,5.1\\
\textcolor{white}{1}\\
\hline
\textcolor{white}{1}\\
O-56 & 1.7 &$^{1}$3$\,\pm\,1$& $-$ & 2.8\,-\,3.5 (2.2\,-\,2.7) & 2.5\,-\,3.5 (1.7\,-\,2.3) & $\simeq$\,9.2$-$28 (0.8$-$1.7)& 1.3\,-\,3.3\\
\textcolor{white}{1}\\
\hline
\textcolor{white}{1}\\
W-4 & 1.7 &$^{5}$$5.2^{+3.8}_{-3.2}$ & $all$&  2.7\,-\,2.9 (1.9\,-\,1.9) & 2.6\,-\,3.0 ($\simeq\,$1.5) & $\simeq$\,1.5$-$2.0 (0.05$-$0.06)& 4.9\,-\,6.3 ($\simeq\,$5.0\,-\,6.2)\\
\textcolor{white}{1}\\
\hline
\textcolor{white}{1}\\
W-19 &1.2 & $^{6}$11.5$^{+2.8}_{-2.7}$ &  $10^{6}$&3.2\,-\,3.3 ($-$) & 3.5\,-\,3.7 ($-$) & 34\,-\,43 ($-$) & 12.3\,-\,13.1 ($-$)\\
       &       					             & & $10^{7}$& 3.3 (1.8\,-\,1.9) & 3.6 (1.4\,-\,1.6) & 36\,-\,38 ($\simeq$\,0.6$-$0.8) & 12.5\,-\,12.6 (12.3\,-\,13.3)\\
              &       					             & & $10^{8}$& 3.3 (1.8) & 3.7 (1.4\,-\,1.5) & 36 ($\simeq$\,0.7$-$0.8) & 12.5 (12.7\,-\,13.3)\\
       &       					             & & $10^{10}$&$-$ (1.8) & $-$ (1.4\,-\,1.5) & $-$($\simeq$\,0.7$-$0.8) & $-$ (12.7\,-\,13.3)\\
\textcolor{white}{1}\\
\hline
\textcolor{white}{1}\\
W-48 & 1.8 & $^{2}$7.9$^{+2.0}_{-1.6}$ &$-$& 2.8\,-\,2.9 (2.5\,-\,2.6) & 4.1\,-\,4.2 (3.4\,-\,3.6) & 8.5\,-\,10 ( $\sim\,$1.6$-$2.0) & 4.5\,-\,4.6\\
\textcolor{white}{1}\\
\hline
\textcolor{white}{1}\\
W-52 & 1.6  & $\textgreater\,0.5$ & $all$& 1.7\,-\,2.0 ($-$) & 1.8\,-\,2.4 ($-$) & $\sim$\,0.07$-$0.2 ($-$) & 1.5\,-\,7.1 ($-$)\\
%	   &         &                                               & $10^{7}$& 1.6\,-\,1.8 (1.6\,-\,1.6) & 1.7\,-\,2.0 (1.7\,-\,1.7) & $\sim\,0.1$ ($\sim\,0.001$\,-\,0.01) & 0.004\,-\,2.2 (0.03\,-\,0.1)\\
%	 &         &                                               & $10^{8}$&  1.6\,-\,1.8 (1.6\,-\,1.6) & 1.7\,-\,2.0 (1.7\,-\,1.7) & $\sim\,0.1$ ($\sim\,0.001$\,-\,0.01) & 0.004\,-\,2.2 (0.02\,-\,0.1)\\
%	 &         &                                               & $10^{10}$& 1.6\,-\,1.8 (1.6\,-\,1.6) & 1.7\,-\,2.0 (1.7\,-\,1.7) & $\sim\,0.1$ ($\sim\,0.001$\,-\,0.01) & 0.004\,-\,2.2 (0.03\,-\,0.1)\\
\textcolor{white}{1}\\
\hline
\textcolor{white}{1}\\
W-78 & 1.8 &  $^{3}$1.37$^{+1.91}_{-0.78}$& $-$&1.9\,-\,2.1  & 2.5\,-\,2.8  & 5.5\,-\,19 & 2.8\\
\textcolor{white}{1}\\
\hline
\textcolor{white}{1}\\
W-103& 1.3 & $^{4}$3\,-\,5& $-$ & 3.1\,-\,3.3 (2.4\,-\,3.0) & 3.3\,-\,3.7 (2.2\,-\,3.1) & 64\,$-$\,116 (5.2\,$-$\,7.9) & 2.8\,-\,3.4 (2.7\,-\,3.4)
%WTS-2& $-$ & 1.5 & & 1.2\,-\,2.6 & 0.7\,-\,2.3 & 0.1\,-\,7.2&1.4\,-\,1.6 & 0.9\,-\,1.1 & 4.1\,-\,6.6\\
\enddata
\tablecomments{See Table~\ref{Tab:systemsParams} for system names. The parameter $t_{*}$ is the stellar age, while $T_{\rm shift}$ is the transit arrival time shift (see text). The subscripts ``lit'' and ``mod'' refer to the literature and our modeling, respectively. The subscripts ``in'' and ``pr'' refer to initial (at the ZAMS) and present values, respectively. For $a_{\rm in}$, $P_{\rm orb, in}$, $T_{\rm shift}$, and $t_{*, mod}$ we list outside of and in parenthesis the parameters derived using the {\it Zahn} and {\it GN} prescription for tides, respectively (\S~\ref{The Orbital Evolution}). If there is no parenthesis, the two numbers agree (e.g., in W-78 the efficiency of tides is never reduced).}
\tablerefs{Following \href{http://exoplanet.eu/catalog/}{exoplanet.eu};
$(1)$: \citealt{Sasselov2003};
$(2)$: \citealt{Enoch+11b};
$(3)$: \citealt{Smalley+12} and references therein;
$(4)$: \citealt{Gillon+14};
$(5)$: \citealt{Gillon+09};
$(6)$: \citealt{Adams+11};
$(7)$: \citealt{Hebrard+13}.}
\label{Tab:orbiEvolResults}
\end{deluxetable}
%%%%%%%%%%%%%%%%%%%%%%%%%%%%%%%%%%%%%%%%%%%%%%%%

\end{document}